# A Sparse Johnson–Lindenstrauss Transform

Anirban Dasgupta          Ravi Kumar          Tamás Sarlós
Yahoo! Research
701 First Avenue
Sunnyvale, CA 94089.
{anirban, ravikumar, stamas}@yahoo-inc.com

## ABSTRACT

Dimension reduction is a key algorithmic tool with many applications including nearest-neighbor search, compressed sensing and linear algebra in the streaming model. In this work we obtain a *sparse* version of the fundamental tool in dimension reduction — the Johnson–Lindenstrauss transform. Using hashing and local densification, we construct a sparse projection matrix with just $\tilde{O}(\frac{1}{\epsilon})$ non-zero entries per column. We also show a matching lower bound on the sparsity for a large class of projection matrices. Our bounds are somewhat surprising, given the known lower bounds of $\Omega(\frac{1}{\epsilon^2})$ both on the number of rows of any projection matrix and on the sparsity of projection matrices generated by natural constructions.

Using this, we achieve an $\tilde{O}(\frac{1}{\epsilon})$ update time per non-zero element for a $(1 \pm \epsilon)$-approximate projection, thereby substantially outperforming the $\tilde{O}(\frac{1}{\epsilon^2})$ update time required by prior approaches. A variant of our method offers the same guarantees for sparse vectors, yet its $\tilde{O}(d)$ worst case running time matches the best approach of Ailon and Liberty.

**Categories and Subject Descriptors.** F.2.0 [**Theory of Computation**]: Analysis of Algorithms and Problem Complexity—*General*; G.3 [**Mathematics of Computing**]: Probability and Statistics—*Probabilistic algorithms*

## 1. INTRODUCTION

Dimension reduction is a fundamental primitive with many algorithmic applications including nearest-neighbor search [2, 19], compressed sensing [11], data stream computations [5], computational geometry [13], numerical linear algebra [14, 17, 26, 28], machine learning [8, 33], graph sparsification [30], and more; see the monograph [32] for further applications. The seminal random projection method of Johnson and Lindenstrauss [20] consists of just multiplying the input vector by a suitably sampled random projection matrix — $n$ vectors in $d$-dimensional space can be mapped into an $O(\frac{1}{\epsilon^2} \log n)$-dimensional subspace such that the length of each vector is distorted by at most $(1 \pm \epsilon)$. This simple and elegant method has the following desirable properties: (i) it is linear, (ii) it is oblivious to the input, (iii) it works with high probability for a given set of input points, and (iv) the target dimension is *independent* of $d$.

Given its algorithmic importance, much effort has been devoted to speeding up the mapping. One line of work achieves this goal by making the projection matrix sparse, and hence its multiplication with the input vectors faster. Sparsity is typically achieved by independently setting each matrix entry to zero with a certain probability [1, 2, 23]. There is however a limit on the extent of sparsity achievable by this approach: a result of Matousek [23, Theorem 4.1] states that such matrices need to contain $\tilde{\Omega}(\frac{\alpha^2}{\epsilon^2})$ non-zeroes in expectation per column, if they were to preserve the length of a unit vector with infinity norm at most $\alpha$.

**Our results.** We obtain a sparse random projection matrix of size $k \times d$ that has $O(\frac{1}{\epsilon} \log^2(\frac{k}{\delta}) \log(\frac{1}{\delta}))$ non-zero entries per column, where $k = O(\frac{1}{\epsilon^2} \log(\frac{1}{\delta}))$. This is the *first* construction with $o(\frac{1}{\epsilon^2})$ non-zero entries in the projection matrix. (For our results to be improvements, we need to assume that $\log^2(\frac{k}{\delta}) = o(\frac{1}{\epsilon})$. Our analysis, however, does not need this assumption.)

A highlight of our approach is to construct the projection matrix itself with care. Instead of using independent random variables, as is typically done, we construct it out of a hash function that entails some dependency among the entries. This construction is implicit in the work of Langford et al. [21] and Weinberger et al. [33], where it played a role mostly as a practical heuristic. The hash-based construction introduces new technical difficulties, but ensures that we have exactly a fixed number of non-zero entries in each column, thereby relaxing the requirements on the density of input vectors.

Specifically, whereas prior work requires that for a unit vector $x$, $\|x\|_\infty = O(\epsilon)$, for a constant number of expected entries per column of the projection matrix, we only need $\|x\|_\infty = O(\sqrt{\epsilon})$. In order to achieve this level of densification, we can use a simple replication technique on $x$ [33].

To manage the technical difficulties that arise from the dependencies, we show that the contribution from each hash bucket is bounded, and that the total amount of noise arising from the collisions in each hash bucket is small. The reduction in overall variance comes from the fact that each dimension is mapped to *exactly one* hash bucket, and the lack of self-collisions (which would be present if the entries in the matrix were i.i.d.) leads to a reduction in the variance of the cross-product error. There are several subtleties in analyzing this, in particular, the errors from different hash buckets being correlated. We handle this by an application of the FKG inequality on the product of the moment generating function of the random variables capturing the errors. This helps us in obtaining a concentration on the sum of the errors. Our choice of $\pm 1$ random variables (instead of Gaussian random variables[1]) plays a critical role in making our proofs work.

**Implications for sparse vectors.** The resulting running time for an input vector $x$ having $n_{\text{nz}}(x)$ non-zeros is $\tilde{O}\left(\frac{n_{\text{nz}}(x)}{\epsilon}\right)$ — better than the running time obtained by [22,23] for sparse vectors in terms of the sparsity ratio $\frac{n_{\text{nz}}(x)}{d}$ as well as by the factor $\frac{1}{\epsilon}$. Furthermore, using a block-Hadamard based preconditioner, instead of a global Hadamard transform, we can actually ensure that our running time for all vectors is $\tilde{O}(\min(\frac{n_{\text{nz}}}{\epsilon}, d))$, which is once again an improvement over existing results. The qualitative difference in the running times is starker in the turnstile model of streaming. Since the updates in the stream come as $(i, v_i)$, updating any sketch that requires computing a global Hadamard transform is very expensive, taking $\tilde{O}(d)$ time per update. Our update time, on the other hand, is only $\tilde{O}(\frac{1}{\epsilon})$ per entry.

Our technique speeds up nearest-neighbor computation for sparse vectors as well. We can use our construction to preprocess the input vectors before applying the algorithm as described in [2, Theorem 3.2]. The effective running time is then $\tilde{O}(\frac{n_{nz}(x)}{\epsilon} + \frac{1}{\epsilon^2}\log n + \frac{1}{\epsilon^3}\log n)$ instead of $O(d \log d + \frac{1}{\epsilon^3}\log n)$. For sparse vectors, this could represent a significant improvement.

**Related work.** Since the original Johnson–Lindenstrauss result, several authors have shown that the projection matrix could be constructed element-wise using Gaussian or uniform $\pm 1$ variables [1,7,16,19]. Alon showed a lower bound of $\Omega\left(\frac{\log n}{\epsilon^2 \log(\frac{1}{\epsilon})}\right)$ on the target dimensionality [4].

In order to circumvent the sparsity lower bound of Matousek [23], the ingenious Fast Johnson–Lindenstrauss transform (FJLT) of Ailon and Chazelle preconditions the input with a randomized Hadamard transform thereby making it dense, and then applies a sparse projection matrix [2]. The computation of the Hadamard transform (via a fast Hadamard transform), however, forces an $\tilde{O}(d)$ running time *irrespective* of the number of non-zeros in the input vector. This makes it less desirable for sparse input vectors.

Ailon and Liberty [3] showed that the sparse projection matrix in [2] could be replaced by a dense, deterministic, but well-structured code matrix, and improved the running time to $O(d \log k)$ over a wide range of parameters; however, like before, the running time of these methods are unable to take advantage of the sparsity of the input vector. Liberty, Ailon, and Singer [22] proved that there exists projection matrices that are applicable in $O(d)$ time if the input satisfies density conditions that are significantly stricter than those required for hashing. Since hashing works in linear time, our work improves upon these results. Finally we remark that although [3,22] contain a spectral condition derived from Talagrand's inequality that could be applied to our hashing construct[2], but the resulting bound is too weak; it fails to show that hashing improves over even the most basic Johnson–Lindenstrauss transform.

Charikar, Chen, and Farach-Colton [12] introduced the COUNT SKETCH data structure that used hash tables combined with pairwise independent $\pm 1$ random variables for finding the most frequent items in a data stream. Thorup and Zhang [31] observed that this hashing trick could be used to speed up the celebrated AMS sketch [5] for estimating $F_2$; this was also noted by Cormode and Garofalakis [15]. Hashing decreases the update time from $O(\frac{1}{\epsilon^2}\log(\frac{1}{\delta}))$ to $O(\log(\frac{1}{\delta}))$. These estimators, however, are non-linear: they return the median of estimates obtained from $O(\log(\frac{1}{\delta}))$ independent hash functions, which makes them less desirable for some applications. Our results essentially show that by increasing the update time to $\tilde{O}(\frac{1}{\epsilon}\log(\frac{1}{\delta}))$, the median could be replaced by an average.

Lastly, we note that random projection using hashing has found practical applications in machine learning [21,29,33]. In particular, the densification by replication was suggested by Weinberger et al. [33]. Although they claim a concentration bound for hashing-based dimensionality reduction, unfortunately, their claim is false due to an error in the application of Talagrand's inequality.

## 2. MAIN RESULTS

Let $k = \frac{12}{\epsilon^2}\log(\frac{1}{\delta})$ and $c = \frac{16}{\epsilon}\log\left(\frac{1}{\delta}\right)\log^2\left(\frac{k}{\delta}\right)$. Let $r = \{r_j\}_{j \in [cd]}$ be a set of i.i.d. random variables such that for each $j \in [cd]$, $\Pr[r_j = 1] = \Pr[r_j = -1] = 1/2$. Let $\delta_{\alpha\beta} = 1$ iff $\alpha = \beta$ and zero otherwise. Let $n_{\text{nz}}(x)$ denote the number of non-zero entries in vector $x$.

Let $h' : [cd] \to [k]$ be a hash function chosen uniformly at random and let $H' \in \{0, \pm 1\}^{k \times cd}$ be defined as $H'_{ij} = \delta_{ih'(j)}r_j$. Let the *pre-conditioner* $P \in \{0, \pm 1\}^{cd \times d}$ be defined as

$$P_{ij} = \begin{cases} \frac{1}{\sqrt{c}} & \text{for } (j-1)c + 1 \leq i \leq jc, \\ 0 & \text{otherwise.} \end{cases}$$

Let $\Phi = H'P$.

**Theorem 1** *For any given vector $x \in \mathbb{R}^d$, with probability $1 - 4\delta$, $\Phi$ satisfies the following property:*

$$(1-\epsilon)\|x\|_2^2 \leq \|\Phi x\|_2^2 \leq (1+\epsilon)\|x\|_2^2. \tag{1}$$

*The time required to compute $\Phi x$ is $O\left(\frac{1}{\epsilon}\log^2(\frac{k}{\delta})\log(\frac{1}{\delta})\right) \cdot n_{\text{nz}}(x)$.*

This is easily implied by the following. Let $h : [d] \to [k]$ be a hash function chosen uniformly at random. Let $H \in \{0, \pm 1\}^{k \times d}$ be defined as $H_{ij} = \delta_{ih(j)}r_j$; note that the matrix $H$ has only $d$ non-zero entries, exactly one per column.

**Theorem 2** *For any given vector $x \in \mathbb{R}^d$ such that $\|x\|_\infty \leq \frac{1}{\sqrt{c}}$, for $\epsilon < 1$ and $\delta < \frac{1}{10}$, with probability $1 - 3\delta$, $H$ satisfies the following property:*

$$(1-\epsilon)\|x\|_2^2 \leq \|Hx\|_2^2 \leq (1+\epsilon)\|x\|_2^2.$$

For dense vectors, Theorem 1 gives a run-time of $O(\frac{d}{\epsilon}\log^3(\frac{1}{\epsilon\delta}))$; this, for a small enough $\epsilon$, could be significantly worse than the running time obtained by Ailon and Liberty in [3] and Matousek in [23]. However, we can modify the construction of the preconditioner so that we guarantee a running time of $O(d \log c \log \log c)$ for *all* vectors. Our new preconditioner is based on the randomized Hadamard construction by Ailon et al. [2,3].

**Theorem 3** *Let $d > 6c\log(\frac{3c}{\delta})$. There exists a preconditioner $G \in \Re^{d \times d}$ such that for any input vector $x \in \mathbb{R}^d$, with probability $1 - 4\delta$,*

$$(1-\epsilon)\|x\|_2^2 \leq \|(HG)x\|_2^2 \leq (1+\epsilon)\|x\|_2^2.$$

*The time required to compute $(HG)x$ is given by*

$$O\left(\min\left(\frac{n_{\text{nz}}(x)}{\epsilon}\log^4\left(\frac{1}{\epsilon\delta}\right), d\right)\log\left(\frac{1}{\epsilon\delta}\right)\right).$$

---
[1] In fact, we need an average of $\frac{1}{\epsilon^2}$ Gaussians to get a $(1 \pm \epsilon)$-approximation.
[2] It is not hard to see that $\sigma$ of [22] equals to $\max\{\sigma_i\}$ studied in Lemma 6.

# 3. PROOF OF THEOREM 2

## 3.1 Preliminaries

Without loss of generality, we can assume $\|x\|_2 = 1$. Let $Y_i = \sum_j H_{ij} x_j = \sum_j \delta_{ih(j)} r_j x_j$, and let $\sigma_i^2 = \mathrm{E}_r[Y_i^2]$, where $\mathrm{E}_r$ is the expectation taken with respect to the random variables $r = \{r_j\}$. Thus,

$$\sigma_i^2 = \mathrm{E}_r[Y_i^2] = \mathrm{E}_r\left[\left(\sum_{j\in[d]} \delta_{ih(j)} r_j x_j\right)^2\right] = \sum_{j\in[d]} \delta_{ih(j)} x_j^2,$$

since the cross-product terms cancel out by the independence i.e., $\mathrm{E}_r[r_j r_{j'}] = 0$ for $j \neq j'$. Let $Z_i = Y_i^2 - \mathrm{E}_r[Y_i^2] = Y_i^2 - \sigma_i^2$.

The outline of the proof is as follows. We need to prove that $\sum_i Y_i^2$ is concentrated around $\|x\|_2^2 = 1$. Instead of showing concentration of $\sum_i Y_i^2$, we will show that $\sum_i Z_i$ is concentrated around zero. Indeed, since our hash function guarantees that each coordinate $j \in [d]$ is mapped to one and exactly one hash bucket, we have that $\sum_{i=1}^k \sigma_i^2 = \|x\|_2^2 = 1$. Therefore, $\sum_{i=1}^k Y_i^2 = \sum_i \sigma_i^2 + \sum_i Z_i = 1 + \sum_{i=1}^k Z_i$. Showing that $\sum_i Z_i$ is concentrated around zero is thus enough.

We will utilize the following form of the FKG inequality [6, Theorem 6.2.1].

**Theorem 4 (FKG inequality)** *Let $L$ be a finite distributive lattice and let $\mu : L \to \Re^+$ be a log-supermodular function. Then, for an increasing function $f$ and a decreasing function $g$, we have that*

$$\sum_{x\in L} \mu(x) f(x) g(x) \sum_{x\in L} \mu(x) \leq \sum_{x\in L} \mu(x) f(x) \sum_{x\in L} \mu(x) g(x).$$

## 3.2 Notation

Recall that $k = \frac{12}{\epsilon^2} \log(\frac{1}{\delta})$. Define

$$\alpha = \alpha(k) = \frac{1}{\epsilon \ln(\frac{k}{\delta})}, \quad \sigma_*^2 = \frac{1+\alpha}{k}, \quad \text{and} \quad \theta = \frac{4\sigma_*^2 k}{\delta};$$

we will assume $\alpha \geq 3$. We define the following function as a shorthand to denote the upper bound on conditional expectation of the MGF with respect to the $\{r_j\}$ variables.

$$G(u,t) = 1 + \frac{1}{\theta^2}(\exp(u\theta) - 1 - u\theta)t + \frac{4\delta}{k}.$$

**Definition 5 (Goodness)** *A set $A \subseteq [d]$ is good if $\sum_{j\in A} x_j^2 \leq \sigma_*^2$. The $i$th hash bucket is good if $h^{-1}(i)$ is good, i.e., if $\sigma_i^2 = \sum_{j,h(j)=i} x_j^2 \leq \sigma_*^2$ and the hash function $h$ is good if $h^{-1}(i)$ is good for all $i$.*

For a given $h$, let $\mathcal{G}_i$ denote the event that the $i$th hash bucket is good. Let $\mathcal{G}$ be the event that the hash function $h$ is good. By abusing notation we use $\mathcal{G}$ and $\mathcal{G}_i$ to represent the indicator variables of the corresponding events.

## 3.3 Proof details

Recall that $Z_i = Y_i^2 - \mathrm{E}_r[Y_i^2]$, where $Y_i = \sum_j \delta_{ih(j)} r_j x_j$, i.e.,

$$Z_i = \sum_{j\neq j', j,j'\in[d]} \delta_{ih(j)} \delta_{ih(j')} r_j r_{j'} x_j x_{j'}.$$

Observe that $\mathrm{E}[Z_i] = 0$ and our goal is to show that $\sum_i Z_i$ is concentrated around 0.

Here is an overview of the proof. We first show in Lemma 6 that most $h$ are good. In Lemma 7, we bound the moment generating function (MGF) of the random variable $Z_i$, for a fixed $h$. A usual step at this point would be to remove the effect of the bad choice of the random variables from the MGF by perhaps considering a truncated random variable $\hat{Z}_i = \min(Z_i, M)$. In our case, however, such a construction would introduce a dependence among the $\{r_j\}$ and $h$ variables, which appears to be insurmountable when trying to apply the FKG inequality. We have to instead utilize the notion of goodness of $h$ only in defining the truncated random variable $\hat{Z}_i$. Using the result of Lemma 7, we first get Corollary 8 that gives the expected and the worst-case bounds on the MGF for a good hash function $h$. We utilize these bounds to define $\hat{Z}_i$ in (5). Next, in Lemma 9, we define two set functions $f_s$ and $g_s$ and show that they are monotone, in accordance with the requirements of the FKG inequality (Theorem 4). These functions are then used in Lemma 10 to show that the MGF of $\sum_i Z_i$ can be bound by the product of the individual MGF's $Z_i$. We then bound the probability of an $\epsilon$-deviation for $\sum_i Z_i$ in Theorem 11. Subsequently, we use Theorem 11 to prove Theorems 1 and 2. Section 4 gives the proof of Theorem 3.

**Lemma 6** *If $c = \frac{16}{\epsilon} \log(\frac{1}{\delta}) \log^2(\frac{k}{\delta})$, then $\Pr[\mathcal{G}] \geq 1 - \delta$.*

The proof (Appendix 9.1) is an application of the Bernstein's inequality [24, Theorem 2.7] and utilizes the fact that since $\|x\|_\infty \leq \frac{1}{\sqrt{c}}$, and the hash function is random, with high probability, no $\sigma_i$ can be too large. In essence, this generalizes well-known facts about the maximum load in the balls into bins problem for the weighted case[3].

The following lemma gives a bound on the MGF of the variable $Z_i$ for a fixed $h$. The proof can be found in Appendix 9.2.

**Lemma 7** *If $u < \frac{1}{4\sigma_i^2}$, then for a fixed $h$,*

$$\mathrm{E}_r[\exp(uZ_i)] \leq G(u, \mathrm{E}_r[Z_i^2]). \qquad (2)$$

Lemma 7 leads to the following.

**Corollary 8** *If $0 < u < \frac{1}{4\sigma_*^2}$, then the expectation of the MGF can be bounded as*

$$\mathrm{E}_{h,r}[\exp(uZ_i) \mid \mathcal{G}] \leq G(u, \frac{1}{k^2}). \qquad (3)$$

*Similarly,*

$$\max_{h\in\mathcal{G}} \mathrm{E}_r[\exp(uZ_i)] \leq G(u, \sigma_*^4). \qquad (4)$$

PROOF. By taking expectation over $h$ and using

$$\mathrm{E}_h[\mathrm{E}_r[Z_i^2 \mid \mathcal{G}]] \leq 2\mathrm{E}[Z_i^2] \leq \frac{2}{k^2},$$

we have that

$$\mathrm{E}_{r,h}[\exp(uZ_i) \mid \mathcal{G}] \leq 1 + \frac{2}{k^2\theta^2}(\exp(u\theta) - 1 - u\theta) + \frac{4\delta}{k}$$
$$\leq \exp\left(\frac{2}{k^2\theta^2}(\exp(u\theta) - 1 - u\theta) + \frac{4\delta}{k}\right).$$

---
[3]Sanders [27] contains a proof of the expected load for the weighted ball-and-bins problem, but does not contain a proof of the high probability statement.

The upper bound on $\mathrm{E}_r[\exp(uZ_i) \mid \mathcal{G}]$ is given by

$$\max_{h \in \mathcal{G}} \mathrm{E}_r[\exp(uZ_i)]$$
$$\leq 1 + \frac{1}{\theta^2}(\exp(u\theta) - 1 - u\theta) \max_h \mathrm{E}_r[Z_i^2 \mid \mathcal{G}] + \frac{4\delta}{k}$$
$$\leq 1 + \frac{1}{\theta^2}(\exp(u\theta) - 1 - u\theta)\sigma_*^4 + \frac{4\delta}{k},$$

where we use $\mathrm{E}_r[Z_i^2] < \sigma_*^4$. □

Next, we have to handle the fact that the $Z_i$ variables are not independent. Yet, intuitively, since $Z_i$ is roughly related to the cross-product of the set of entries $x_j$ that map into the $i$th hash bucket, conditioned on the fact that one of the $Z_i$ variables has achieved a large value, the probability that another $Z_{i'}$ is also large decreases. In fact, we show that we can apply the FKG inequality (Theorem 4) on the MGF of the $Z_i$ random variables. Note that this situation is more involved that the simple negative dependence obtained on a set of random variables by conditioning their sum to be a constant — we cannot make such claims on $\sum_i Z_i$. For all $i = 1, \ldots, k$ let us define

$$\hat{Z}_i = \begin{cases} Z_i & \text{if } \mathcal{G}_i, \\ \frac{1}{u}\log G(u, \sigma_*^4) & \text{else.} \end{cases} \quad (5)$$

We first need the following lemma in preparation for the application of the FKG inequality (Theorem 4).

**Lemma 9** *For $1 \leq s \leq k$, $u < \frac{1}{4\sigma_*^2}$ and $A \subseteq [d]$, let us define*

$$f_s(A) = \mathrm{E}_r\left[\exp\left(u\hat{Z}_s\right) \mid h^{-1}(s) = A\right] \text{ and }$$
$$g_s(A) = \mathrm{E}_{h,r}\left[\exp\left(u\sum_{i=1}^{s-1}\hat{Z}_i\right) \mid h^{-1}(s) = A\right].$$

*Then $f_s$ is an increasing and $g_s$ is a decreasing set function.*

PROOF. First we prove that $f_s$ is increasing by showing that for all $A \subseteq [d]$ and for all $a \in [d] \setminus A$, it holds that $f_s(A \cup \{a\}) \geq f_s(A)$.

Observe that if $h^{-1}(s)$ is good (i.e., if $\mathcal{G}_s$ holds), then by Corollary 8, we have $\mathrm{E}_r[\exp(uZ_s)] \leq G(u, \sigma_*^4)$. Thus for all $h$ and $s$, it holds from (5) that

$$\mathrm{E}_r[\exp(u\hat{Z}_s)] \leq G(u, \sigma_*^4). \quad (6)$$

There are two cases to consider. Suppose $A \cup \{a\}$ is bad. Then, $\hat{Z}_s = \frac{1}{u}\log G(u, \sigma_*^4)$ and hence $f_s(A \cup \{a\}) = G(u, \sigma_*^4) \geq f_s(A)$ from (6).

Suppose $A \cup \{a\}$ is good. Now, let us define

$$V_A = \sum_{j,g \in A, j \neq g} r_j r_g x_j x_g \quad \text{and} \quad W_A = x_a \sum_{j \in A} x_j r_j.$$

Also note that if $h^{-1}(s) = A \cup \{a\}$ and the $s$th bucket is good, then $\hat{Z}_s = Z_s = V_A + r_a W_A$ holds. Therefore we have that

$$f_s(A \cup \{a\}) = \mathrm{E}_r\left[\exp\left(u\hat{Z}_s\right) \mid h^{-1}(s) = A \cup \{a\}\right]$$
$$= \mathrm{E}_r\left[\exp\left(uV_A + u \cdot r_a W_A\right) \mid h^{-1}(s) = A \cup \{a\}\right]$$
$$= \mathrm{E}_r[\mathrm{E}_r\left[\exp\left(uV_A + u \cdot r_a W_A\right) \mid h^{-1}(s) = A \cup \{a\}, \{r_j\}_{j \in A}\right]$$
$$\quad \mid h^{-1}(s) = A \cup \{a\}]$$
$$\geq \mathrm{E}_r[\exp\left(\mathrm{E}_r\left[uV_A + u \cdot r_a W_A \mid h^{-1}(s) = A \cup \{a\}, \{r_j\}_{j \in A}\right]\right)$$
$$\quad \mid h^{-1}(s) = A \cup \{a\}].$$
(By Jensen's inequality, $\mathrm{E}[\exp(x)] \geq \exp(\mathrm{E}[x])$)
$$\stackrel{(a)}{=} \mathrm{E}_r\left[\exp\left(uV_A\right) \mid h^{-1}(s) = A \cup \{a\}\right]$$
$$\stackrel{(b)}{=} \mathrm{E}_r\left[\exp\left(uV_A\right) \mid h^{-1}(s) = A\right]$$
$$\stackrel{(c)}{=} f_s(A).$$

Here, (a) follows since only $r_a$ is random in the inner expectation and

$$\mathrm{E}_r\left[uV_A + u \cdot r_a W_A \mid h^{-1}(s) = A \cup \{a\}, \{r_j\}_{j \in A}\right] = uV_A.$$

And, (b) follows since $a \notin A$ and $V_A$ does not depend on $h(a)$ by the independence of the values of $r$ and $h$. Finally, (c) follows since if $A \cup \{a\}$ is good then so is $A$; therefore if $h^{-1}(s) = A$, then we have that $\hat{Z}_s = Z_s = V_A$. The proof that $f_s$ is increasing is complete.

The proof of $g_s$ being a decreasing function is similar, and can be found in Appendix 9.3. □

Given our construction of the two functions, $f_s$ and $g_s$, we can now proceed to apply the FKG inequality (Theorem 4) to show that the MGF of the random variable $\sum_{i=1}^k \hat{Z}_i$ is bounded by the product of the MGF's of each $\hat{Z}_i$ variable.

**Lemma 10** *It holds that*

$$\mathrm{E}\left[\exp\left(u\sum_{i=1}^k \hat{Z}_i\right)\right] \leq \prod_{i=1}^k \mathrm{E}\left[\exp\left(u\hat{Z}_i\right)\right],$$

*where the expectation is taken over both $h$ and $r = \{r_j\}$.*

PROOF. For all $1 \leq s \leq k$, we prove

$$\mathrm{E}\left[\exp\left(u\sum_{i=1}^s \hat{Z}_i\right)\right] \leq \prod_{i=1}^s \mathrm{E}\left[\exp\left(u\hat{Z}_i\right)\right], \quad (7)$$

by induction on $s$. The base case of $s = 1$ is obvious.

Now assume that the inductive hypothesis (7) holds for $s - 1$. For all $A \subseteq [d]$ let us define

$$\mu_s(A) = \Pr\left[h^{-1}(s) = A\right] = \prod_{j \in A} \Pr[h(j) = s] \prod_{j \notin A} \Pr[h(j) \neq s].$$

It is easy to check that $\mu_s$ is a log-supermodular measure[4] over the subsets of $[d]$. Recalling the definition of the increasing function $f_s$ and the decreasing function $g_s$ from Lemma 9 it follows from the FKG inequality (Theorem 4) that

$$\mathrm{E}_{\mu_s}[f_s g_s] \leq \mathrm{E}_{\mu_s}[f_s] \mathrm{E}_{\mu_s}[g_s].$$

---
[4] See [6, Section 6.2, page 87] for a precise definition and proof of this fact.

Furthermore, observe that for any random variable $X$ we have

$$\mathrm{E}_{\mu_s}\left[\mathrm{E}\left[X \mid h^{-1}(s) = A\right]\right] =$$
$$\sum_{A \subseteq [d]} \Pr\left[h^{-1}(s) = A\right] \mathrm{E}\left[X \mid h^{-1}(s) = A\right] = \mathrm{E}[X],$$

and consequently,

$$\mathrm{E}[\exp(u\sum_{i=1}^{s-1}\hat{Z}_i)\exp(u\hat{Z}_s)] \leq \mathrm{E}[\exp(u\sum_{i=1}^{s-1}\hat{Z}_i)]\mathrm{E}[\exp(u\hat{Z}_s)].$$

Combining the latter with the induction hypothesis for $s-1$ concludes the proof. □

**Theorem 11** *For the variables $Z_i$ we have*

(i)   $\Pr\left[\sum_i Z_i \geq \epsilon\right] \leq \exp\left(\dfrac{-3k\epsilon^2}{4(3 + (1+\alpha)\epsilon \ln(\frac{k}{\delta}))} + 4\delta\right) + \delta,$

(ii)   $\Pr\left[\sum_i Z_i < -\epsilon\right] < \exp\left(-\dfrac{\epsilon^2 k}{12}\right) + \delta.$

The proof of Theorem 11 involves a standard but tedious calculation that is similar to one done by Matousek [23]. The proof can be found in Appendix 9.4. Finally, we are ready to prove the main result.

PROOF. (**of Theorem 2**). Recall that $Y_i = \sum_j H_{ij}x_j$, thus $\|Hx\|_2^2 = \sum_i Y_i^2$. Also recall that $\sigma_i^2 = \mathrm{E}_r[Y_i^2]$. Thus, $\sum_{i=1}^k \sigma_i^2 = \|x\|_2^2 = 1$. Therefore, $\sum_{i=1}^k Y_i^2 = \sum_i \sigma_i^2 + \sum_i Z_i = 1 + \sum_{i=1}^k Z_i$. Recall that $k = \frac{12}{\epsilon^2}\log(\frac{1}{\delta})$, and $\alpha = \frac{1}{\epsilon \log(\frac{k}{\delta})}$. Plugging these values in Theorem 11(i), we have $\sum_i Z_i > \epsilon$, with probability at most $\exp(-\frac{9}{5}\ln(\frac{1}{\delta}) + 4\delta) + \delta < 2\delta$, for $\delta < \frac{1}{10}$. Similarly, from Theorem 11(ii), we have $\sum_i Z_i < -\epsilon$ with probability at most $2\delta$. Putting them together, with probability at least $1 - 4\delta$, $|\sum_i Y_i^2 - 1| = |\sum_i Z_i| < \epsilon$, and hence $\|\|Hx\|_2^2 - \|x\|_2^2| < \epsilon\|x\|_2^2$. □

PROOF. (**of Theorem 1**). Theorem 1 easily follows from Theorem 2 by noting that if $y = Px$, then $\|y\|_2 = 1$ and $\|y\|_\infty \leq \frac{1}{\sqrt{c}}$. The running time is obtained as computing both $y = Px$ and $Hy$ requires $O(c \cdot n_{nz}(x))$ time. □

## 4. PROOF OF THEOREM 3

**Definition 12 (Randomized Hadamard matrix [2].)** *Construct the $m \times m$ Hadamard matrix $F$ as $F_{ij} = m^{-1/2}(-1)^{\langle i-1, j-1 \rangle}$ and the diagonal matrix $D$ by choosing each $D_{ii}$ independently from $\{-1, 1\}$ with probability $1/2$ for each value. The matrix $A = FD$ is defined to be an $m \times m$ randomized Hadamard matrix.*

Using multiple small copies the randomized Hadamard matrix, we create the following preconditioner. Without loss of generality, we assume that $\frac{d}{b}$ is an integer, for the given value of $b$. We note that [3] also contains a similar construct; here we present a more straightforward analysis using a different vector norm.

**Lemma 13** *Let $x \in \Re^d$, $\|x\| = 1$, and $1 > \delta > 0$, and $c \geq 1$. Define $b = 6c\log(\frac{3c}{\delta})$ and assume $b \leq d$. Construct $G \in \Re^{d \times d}$ to be a random block-diagonal matrix, where each of the $d/b$ diagonal blocks of $G$ consist of an independent copy of a $b \times b$ randomized Hadamard matrix. Then we have that*

$$\Pr\left[\|Gx\|_\infty \geq \frac{1}{\sqrt{c}}\right] \leq \delta.$$

PROOF. If $A$ is $b \times b$ randomized Hadamard matrix, then for any $b$-dimensional vector $z$ with $\|z\|_2 = 1$ it holds that $\|Az\|_2 = 1$. Using a Chernoff-type argument Ailon and Chazelle [2] showed

$$\Pr[\|Az\|_\infty \geq s] \leq 2b\exp\left(-\frac{s^2 b}{2}\right). \quad (8)$$

holds as well. Observe that the previous inequality trivially holds for $\|z\|_2 \leq 1$ as well. Let $y = Gx$, and $G_j$ denote the $j$th diagonal block of $G$, and partition $x$ and $y$ into $\frac{d}{b}$ blocks $x_j$ and define $y_j = G_j x_j$. Now for a block $j$, if $\|x_j\|_2 \leq \frac{1}{\sqrt{c}}$, then $\|y_j\|_\infty \leq \|y_j\|_2 \leq \frac{1}{\sqrt{c}}$ holds as well, since $G_j$ is an isometry. Since $x$ is unit length, there could be at most $c$ blocks $j$ such that $\|x_j\|_2 \geq \frac{1}{\sqrt{c}}$. Thus setting $s$ to $\frac{1}{\sqrt{c}}$ in (8) and taking the union bound over these at most $c$ blocks, we have that

$$\Pr\left[\|Gx\|_\infty \geq \frac{1}{\sqrt{c}}\right] \leq 2bc\exp\left(-\frac{b}{2c}\right) = \frac{12c^2\log(\frac{c}{\delta})\delta^3}{27c^3} \leq \delta,$$

establishing the claim. □

Using the block-Hadamard preconditioner, we are ready to prove Theorem 3. The $\epsilon$-approximation guarantee of the projection matrix $\Phi$ follows trivially from the statements of Theorem 2 and of Lemma 13.

In order to bound the running time, let $n_{\mathrm{nzb}}(x)$ denote the number of blocks that have non-zero coordinates in $x$. Then the running time of the block-Hadamard based hashing is $O(n_{\mathrm{nzb}}(x) \cdot b \log b + n_{\mathrm{nzb}}(x) \cdot b)$. Now,

$$n_{\mathrm{nzb}}(x) \cdot b \log b \leq \min(n_{\mathrm{nz}}(x)b, d) \log b$$
$$= O(\min(n_{\mathrm{nz}}(x)c\log(\frac{c}{\delta}), d)\log(\frac{c}{\delta}))$$

Now, $c\log(\frac{c}{\delta}) = O(\frac{1}{\epsilon}\log(\frac{1}{\delta})\log^2\frac{k}{\delta}\log(\frac{1}{\epsilon\delta}))$. Hence the final running time is

$$O\left(\min\left(\frac{n_{\mathrm{nz}}(x)}{\epsilon}\log\left(\frac{1}{\delta}\right)\log^2\left(\frac{k}{\delta}\right)\log\left(\frac{1}{\epsilon\delta}\right), d\right)\log\left(\frac{1}{\epsilon\delta}\right)\right).$$

Note that if $\delta$ is not too small then the running time of Theorem 3 is comparable to the best existing methods for dense vectors [3] yet it is much faster for sparse vectors. We remark that the localized Hadamard preconditioner presented in this section could also be combined with suitably sparse random matrices from [23] by making $b$ larger, approximately equal to $k$. This variant would reproduce the results of [3], but it fails to show any improvement for sparse vectors over the naive construction as the running time would be $\tilde{\Omega}(\frac{1}{\epsilon^2})$ per non-zero element.

## 5. A LOWER BOUND

A random matrix $\Phi$ is said to have the *JL property* if for every vector $x$, $\Phi x$ satisfies (1) with probability $1 - \delta$ over the choice of $\Phi$.

We show a lower bound on the sparsity for a class of constructions of matrices with the JL property. The construction of the matrix is modeled as a two stage process: first, the set of indices that have non-zero entries is chosen, and then each column is chosen independently random. Note that we do not assume that the random variables are independent within a column.

The lower bound argument of Matousek [23] shows that if the set of non-zero indices in the first stage is chosen by independent coin tosses and if the random variables in the second stage are independent (scaled) $\pm 1$ with equal probability, in expectation, then

$\tilde{\Omega}(\frac{\|x\|_\infty^2}{\epsilon^2})$ non-zero entries per column are needed to guarantee that the resulting matrix has the JL property.

We show a lower bound on the sparsity for the case when the non-zero indices are chosen arbitrarily. As mentioned earlier, if the random variables in the second stage are $N(0,1)$, then it is easy to obtain a lower bound of $\tilde{\Omega}(\frac{1}{\epsilon^2})$ on the number of non-zero entries per column: indeed, the lower bound follows since $\tilde{\Omega}(\frac{1}{\epsilon^2})$ such random variables are needed so that their sum is $(1 \pm \epsilon)$, w.h.p.

Under mild technical conditions on the random variables, we can prove the following lower bound stated in Theorem 14. It is easy to see that the conditions of Theorem 14 are satisfied if the random entries are independent (scaled) $\pm 1$ or when they are generated by the replicated hashing construct of Theorem 1. Thus the upper bound of Theorem 1 is tight with respect to $\epsilon$. The bound on the number of non-zeros per column implies a bound on the worst case update time over all vectors as well.

**Theorem 14** Let $1 \leq c \leq k < d$ be integers and $M$ be an arbitrary, fixed or random, $k \times d$ 0-1 matrix with at most $c$ non-zeroes per column. Let $P$ be a $k \times d$ random matrix of the following form $P_{ij} = \begin{cases} 0 & \text{if } M_{ij} = 0 \\ U_{ij} & \text{if } M_{ij} = 1. \end{cases}$ Here the vector valued $U_{*j}$ random variables are independent and for each $j$ it holds that $\mathrm{E}[\sum_i P_{ij}^2] = 1$ and that $\mathrm{E}[\sum_i P_{ij}^4] = O(\frac{1}{c})$.

Let $0 < \epsilon \leq 1/4$. If $P$ has JL property with probability at least $1 - \frac{o(1)}{d^2}$, then

$$c = \Omega\left(\min\left\{\frac{1}{\epsilon^2}, \frac{\sqrt{\log_k(d)}}{\epsilon}\right\}\right).$$

PROOF. For all $i = 1, \ldots, d$ let $C_i = \{s \in [k] | M_{si} \neq 0\}$ denote the index set of non-zeros in the $i$th column of $P$. Furthermore, let $V = \{e_1, \ldots, e_d\}$, where $e_i$ denotes the $i$th unit vector. For $i \neq j$ we also define $X_{ij} = C_i \cap C_j$ and $S = \sum_{t \in X_{ij}} U_{ti} U_{tj}$. Then we have that

$$\|P(e_i + e_j)\|_2^2 = \|Pe_i\|_2^2 + \|Pe_j\|_2^2 + 2S. \quad (9)$$

Using the fourth moment method [9], we show that $S$ has a large deviation with constant probability unless $c$ is large enough. Towards this goal for all $t \in X_{ij}$ set $Y_t = U_{ti}U_{tj}$ and let $x_{ij} = |X_{ij}|$.

W.l.o.g. we can assume that each column of $M$ contains exactly $c$ non-zeroes and if $M_{ti} = 1$ then $\mathrm{E}[U_{ti}^2] = \frac{1}{c}$ and $\mathrm{E}[U_{ti}^4] = O(\frac{1}{c^2})$ hold as well; otherwise we replace $P$ with a copy of $P$ whose rows are randomly permuted. Furthermore we can also assume that $\mathrm{E}[U_{si}U_{ti}] = 0$ holds as multiplying each row of $P$ with independent uniformly distributed $\pm 1$ random variables does not change (9) or the theorem's conditions. Finally, w.l.o.g. we can assume that for all $s, t_1, t_2, t_3$ where $s \notin \{t_1, t_2, t_3\}$ it holds that $\mathrm{E}[Y_s Y_{t_1} Y_{t_2} Y_{t_3}] = 0$ as multiplying the rows of $P$ with random $\pm 1$ ensures the latter condition as well.

Now observe that $\mathrm{E}[S^2] = \mathrm{E}[\sum_t Y_t^2] + \sum_{s \neq t} \mathrm{E}[Y_s Y_t] = \frac{x_{ij}}{c^2}$ holds, since $\mathrm{E}[Y_t^2] = \mathrm{E}[U_{ti}^2 U_{tj}^2] = \mathrm{E}[U_{ti}^2]\mathrm{E}[U_{tj}^2] = \frac{1}{c^2}$ by the independence of columns. Moreover if $s \neq t$ then we have that $\mathrm{E}[Y_s Y_t] = \mathrm{E}[U_{si}U_{ti}U_{sj}U_{tj}] = \mathrm{E}[U_{si}U_{ti}]\mathrm{E}[U_{sj}U_{tj}] = 0$ by independence again.

Similarly note that

$$\mathrm{E}[S^4] = \mathrm{E}[\sum_t Y_t^4] + \sum_{s \neq t} \mathrm{E}[Y_s^2 Y_t^2] + \sum_{s \notin \{t_1,t_2,t_3\}} \mathrm{E}[Y_s Y_{t_1} Y_{t_2} Y_{t_3}]$$
$$= O\left(\frac{x_{ij}^2}{c^4}\right).$$

By our assumptions it holds that $\mathrm{E}[Y_t^4] = \mathrm{E}[U_{ti}^4]\mathrm{E}[U_{tj}^4] = O(\frac{1}{c^4})$. If $s \neq t$ then $\mathrm{E}[Y_s^2 Y_t^2] = \mathrm{E}[U_{si}^2 U_{ti}^2]\mathrm{E}[U_{sj}^2 U_{tj}^2]$ holds by independence and hence from Hölder's inequality we have that $\mathrm{E}[U_{si}^2 U_{ti}^2] \leq \sqrt{\mathrm{E}[U_{si}^4]\mathrm{E}[U_{ti}^4]}$. Thus it holds that $\mathrm{E}[Y_s^2 Y_t^2] \leq O(\frac{1}{c^4})$. Lastly, recall that that for all $s, t_1, t_2, t_3$ where $s \notin \{t_1, t_2, t_3\}$ we have that $\mathrm{E}[Y_s Y_{t_1} Y_{t_2} Y_{t_3}] = 0$.

Now [9, Theorem 3.5] states that

$$\Pr\left[|S| \geq \frac{1}{2}\sqrt{\mathrm{E}[S^2]}\right] \geq \frac{(3/4)^2}{\frac{\mathrm{E}[S^4]}{\mathrm{E}[S^2]^2} - \frac{7}{16}}.$$

Therefore we have that

$$\Pr\left[|S| \geq \frac{\sqrt{x_{ij}}}{2c}\right] \geq \frac{(3/4)^2}{O(1) - \frac{7}{16}} = \Omega(1). \quad (10)$$

On the other hand, it follows from the assumed JL property of $P$ that with probability $1 - o(1)$, for all $1 \leq i < j \leq d$, we have that $\left|\|P(e_i + e_j)\|_2^2 - 2\right| \leq 2\epsilon$ and that

$$\left|\|P(e_i)\|^2 + \|Pe_j\|_2^2 - 2\right| \leq 2\epsilon.$$

Therefore from combining equality (9) with inequality (10) it follows that $\frac{\sqrt{x_{ij}}}{c} \leq 4\epsilon$ must hold for all $i \neq j$, or equivalently $|C_i \cap C_j| \leq z$ with $z = 16\epsilon^2 c^2$ for all $i \neq j$.

If $z < 1$, then the $C_i$ are pairwise disjoint and therefore $k \geq dc \geq d$, a contradiction. Thus $z \geq 1$, and hence $c \geq \frac{1}{16\epsilon}$ immediately. In what follows we strengthen the latter lower bound for a large range of $d$ and $k$ as claimed.

If $c \geq \frac{1}{32\epsilon^2}$ then the lemma clearly holds as $\Omega\left(\frac{1}{\epsilon^2}\right)$ is the largest of the lower bounds claimed.

Now note that $c \geq 2$. Since if $c = 1$ were to hold, then from $\epsilon < 1/4$ it follows that $z < 1$, which is a contradiction as before. Therefore if $c \leq \frac{1}{32\epsilon^2}$ then all $C_i$'s are distinct as $z + 1 = (c/2)(32\epsilon^2 c) + 1 \leq c/2 + c/2 = c$ holds.

Observe that any $z + 1$ element set is contained in at most one $C_i$. Therefore the number of distinct $C_i$ is at most

$$f(k,c,z+1) = \binom{k}{z+1} / \binom{c}{z+1},$$

a well known fact from block designs and set packing [18]. From the Stirling formula, for all $n \geq 1$, $\sqrt{2\pi n}\left(\frac{n}{e}\right)^n \leq n! \leq 1.1\sqrt{2\pi n}\left(\frac{n}{e}\right)^n$, and it follows that for all $1 \leq y < x$ it holds that $\left(\frac{ex}{y}\right)^y \frac{0.8}{\sqrt{2\pi}}\sqrt{\frac{x-y}{xy}} \leq \binom{x}{y} \leq \left(\frac{ex}{y}\right)^y \frac{1.1}{\sqrt{2\pi}}\sqrt{\frac{x}{y(x-y)}}$. Therefore we have that

$$f(k,c,z+1) \leq \left(\frac{k}{c}\right)^{z+1} 2k \leq k^{z+3}. \quad (11)$$

Now observe that $d \leq f(k,c,z+1)$ as all $C_i$ are distinct. Combining the latter with inequality (11), we have that $\log_k(d) - 3 \leq z$. Recalling that $1 \leq z = 16\epsilon^2 c^2$ concludes the proof. □

Using a replication argument it is easy to see that if a matrix $P$ only has the JL property for vectors $x$ with $\frac{\|x\|_\infty}{\|x\|_2} \leq \alpha$ for some $\alpha$, then under the conditions of Theorem 14 we have that at least one column of $P$ contains $\Omega\left(\alpha^2 \min\left\{\frac{1}{\epsilon^2}, \frac{\sqrt{\log_k(d)}}{\epsilon}\right\}\right)$ non-zeroes.

If the fourth moment of the random entries per column scales with the number of non-zeros per column, the next theorem strengthens the previous claim by bounding the average number of non-zeroes per column. This condition is satisfied, say, if the non-zero entries are independent scaled $\pm 1$ random variables.

**Theorem 15** *Let $0 < \epsilon \leq 1/4$ and $M$ be an arbitrary $k \times d$ 0-1 matrix with $2k^2 < d$. Let $c_j$ denote the number of non-zeroes in the jth column of $M$. Let $P$ be a $k \times d$ random matrix of the following form $P_{ij} = \begin{cases} 0 & \text{if } M_{ij} = 0 \\ U_{ij} & \text{if } M_{ij} = 1 \end{cases}$. Here the vector valued $U_{*j}$ random variables are independent and for each j it holds that $\mathrm{E}[\sum_i P_{ij}^2] = 1$ and $\mathrm{E}[\sum_i P_{ij}^4] = O(1/c_j)$.*

*If $P$ has JL property with probability at least $1 - \frac{o(1)}{d^2}$, then*

$$\sum_{i=1}^{d} \frac{c_i}{d} = \Omega\left(\min\left\{\frac{1}{\epsilon^2}, \frac{\sqrt{\log_k(d)}}{\epsilon}\right\}\right).$$

PROOF. Set

$$s = \Omega\left(\min\left\{\frac{1}{\epsilon^2}, \frac{\sqrt{\log_k(d)}}{\epsilon}\right\}\right).$$

For all $j = 1, \ldots, k$, assemble the columns of $P$ with $c_i = j$ into the $k \times n_j$ matrix $P_j$. For all $j$ if $n_j > k$ then from assumed JL property of $P$ it follows that $P_j$ satisfies the conditions of Theorem 14 with $c = j$ and thus $j \geq s$.

Therefore for all $j < s$ we have that $n_j \leq k$. The number of non-zeroes in $P$ is $\sum_{i=1}^{d} c_i = \sum_{j=1}^{k} n_j j$, which we lower bound as follows

$$\sum_{j=1}^{k} n_j j \geq \sum_{j=s}^{k} n_j s = \left(\sum_{j=1}^{k} n_j - \sum_{j=1}^{s-1} n_j\right) s \geq (d - sk) s$$

$$\geq \left(d - k^2\right) s \geq \frac{d}{2} s. \quad \square$$

## 6. EMBEDDING INTO $\ell_1$

We can show the following result for the case that the target metric is $\ell_1$. The result and the corresponding proof is similar to that of Ailon and Chazelle [2]. We construct the matrix $H$ as follows: $H_{ij} = \delta_{ih(j)} r_j$, where $r_j$ are now drawn i.i.d. random variables $N(0,1)$ instead of being $\pm 1$. We then have the following. Let $\beta_0 = \mathrm{E}[|z|]$ where $z \sim N(0,1)$. By the 2-stability of the normal distribution, $Y_i = \sum_j x_j r_j \delta_{ih(j)} \sim N(0, \sigma_i)$ where $\sigma_i^2 = \sum_{j \in h^{-1}(i)} x_j^2$. Thus, $\mathrm{E}_r[|Y_i|] = \sigma_i \beta_0$.

**Theorem 16** *There exists a constant $\epsilon_0$ such that for all $\epsilon < \epsilon_0$, if $c = k/\epsilon$, and $k = O\left(\frac{1}{\epsilon^2}\log(\frac{1}{\delta})\right)$, $Y = \frac{1}{\beta_0 \sqrt{k}} \sum_i |Y_i|$, we have that $\Pr[|Y - 1| > \epsilon] < \delta$.*

The proof is omitted in this version.

## 7. DISCUSSIONS

The most important open question is resolving the gap between the upper and lower bounds with respect to the error probability. It would be interesting to see whether our claims could be proven more directly using stronger concentration inequalities.

Application of the current result to streaming settings would also require proving the claims for a $k$-wise independent hash-function and $\pm 1$ variables. The chief hurdle in applying the techniques of Clarkson and Woodruff [14] seems to be proving the FKG inequality for the limited independence case. Note that Nisan's pseudorandom number generator construction [25] can be used to derandomize the hash function, but the naive way of doing this increases the update time to $k$. We leave efficient derandomization as an open question.

It is worthwhile to note that the hash-function represents a bipartite expander. In a similar vein, Berinde et al. [10] use an unbalanced expander graph based construction to create matrices with restricted isometry property for sparse signal recovery. Their argument crucially uses two facts — that the error-norm is $\ell_1$, and that the input vector is sparse. It would interesting to investigate possible connections between these results.

**Acknowledgments.** The authors would like to thank Flavio Chierichetti, Edo Liberty, and Alex Smola for helpful discussions. We also thank the anonymous reviewers for feedback and for suggesting future directions.

# 9. APPENDIX

## 9.1 Proof of Lemma 6

PROOF. We show that $\sigma_1^2 \leq \sigma_*^2$, with probability $1 - \delta/k$; the proof will then follow from the union bound.

Define the random variable $X_j = \delta_{1h(j)} x_j^2 - \frac{x_j^2}{k}$. Then, $E_h[X_j] = 0$ and since $\|x\|_\infty < \frac{1}{\sqrt{c}}$, we have $X_j < \frac{1}{c}$. We also have $E_h[X_j^2] = E_h\left[\left(\delta_{1h(j)} x_j^2 - \frac{x_j^2}{k}\right)^2\right] = x_j^4\left(\frac{1}{k} + \frac{1}{k^2} - \frac{2}{k^2}\right) \leq \frac{x_j^2}{c}\left(\frac{1}{k} - \frac{1}{k^2}\right)$, and

$$\sum_j E_h[X_j^2] \leq \sum_j \frac{x_j^2}{c}\left(\frac{1}{k} - \frac{1}{k^2}\right) \leq \frac{1}{c}\left(\frac{1}{k} - \frac{1}{k^2}\right).$$

Also, $\sum_j X_j = \sigma_1^2 - \frac{1}{k}$. Plugging this into the Bernstein's inequality [24, Theorem 2.7],

$$\Pr\left[\sigma_1^2 - \frac{1}{k} > \frac{\alpha}{k}\right] = \Pr\left[\sum_j X_j > \frac{\alpha}{k}\right]$$
$$\leq \exp\left(-\frac{(\alpha/k)^2/2}{\frac{1}{c}\left(\frac{1}{k} - \frac{1}{k^2}\right) + \frac{\alpha}{3ck}}\right) \leq \exp\left(-\frac{c\alpha^2/2}{(k-1) + \alpha k/3}\right)$$
$$\leq \exp\left(-\frac{c\alpha^2/2}{k + \alpha k/3}\right).$$

Since $\alpha \geq 3$,

$$\Pr\left[\sigma_1^2 - \frac{1}{k} > \frac{\alpha}{k}\right] \leq \exp\left(-\frac{c\alpha^2/2}{2k\alpha/3}\right) \leq \exp\left(-\frac{3c\alpha}{4k}\right).$$

Choosing $c = \frac{4k}{3\alpha}\log(\frac{k}{\delta})$, we get the above probability to be smaller than $\delta/k$. Since $\alpha = \frac{1}{\epsilon \log(k/\delta)}$, and $k = \frac{12}{\epsilon^2}\log(1/\delta)$, we have that choosing $c = \frac{16}{\epsilon}\log(\frac{1}{\delta})\log^2(\frac{k}{\delta})$ is sufficient. □

## 9.2 Bounding the MGF's

We first compute the expectation of the MGF for different conditions on the hashing function. We begin by proving Lemma 7.

### 9.2.1 Proof of Lemma 7

PROOF. We have that $Z_i = \sum_{j \neq g \in h^{-1}(i)} r_j r_g x_j x_g$. Hence

$$Z_i = \left(\sum_{j:h(j)=i} x_j r_j\right)^2 - \sum_{j:h(j)=i} x_j^2 = Y_i^2 - \sigma_i^2$$

where $Y_i = \sum_{j:h(j)=i} x_j r_j$. Then,

$$E_r[\exp(uY_i)] = \prod_{j:h(j)=i} E_r[\exp(ur_j x_j)]$$
$$= \prod_{j:h(j)=i}\left(\frac{1}{2}\exp(ux_j) + \frac{1}{2}\exp(-ux_j)\right)$$
$$\leq \prod_{j:h(j)=i} \exp\left(\frac{u^2 x_j^2}{2}\right) \leq \exp\left(\frac{u^2 \sigma_i^2}{2}\right).$$

By the Markov inequality, we get the probability of $Y_i$ being larger than $t$ as

$$\Pr[Y_i > t] \leq \frac{E_r[\exp(uY_i)]}{\exp(ut)} \leq \frac{\exp\left(\frac{u^2 \sigma_i^2}{2}\right)}{\exp(ut)} \leq \exp\left(-\frac{t^2}{2\sigma_i^2}\right),$$

by choosing $u = \frac{t}{\sigma_i^2}$. Note that we do not need to worry about $\sigma_i$ being zero, as then $Y_i = 0$. Then, we bound $E_r[\exp(uZ_i)]$ as follows. Denote $p(t) = \Pr[Z_i = t]$. We first compute the expectation with respect to $r$. For any value of $\theta > 0$, we have

$$E_r[\exp(uZ_i)] = \sum_{t \in (-\infty,\infty)} \exp(ut) p(t)$$
$$\leq \sum_{t \in (-\infty,\theta]} \exp(ut) p(t) + \sum_{t > \theta} \exp(ut) p(t).$$

The first term can be bounded as follows:

$$\sum_{t\in(-\infty,\theta)} \exp(ut)p(t) \leq \sum_{t\in[0,\theta]} \left(1 + ut + \sum_{j=2}^{\infty} \frac{u^j t^j}{j!}\right) p(t)$$

$$\leq \sum_{t\in(-\infty,\theta]} p(t) + \sum_{t\in(-\infty,\theta]} utp(t) + \sum_{t\in(-\infty,\theta]} \sum_{j=2}^{\infty} \frac{u^j t^j}{j!} dF(t)$$

$$\leq \sum_{t\in(-\infty,\infty)} p(t) + \sum_{t\in(-\infty,\infty)} utp(t) + \sum_{t\in(-\infty,\theta]} \sum_{j=2}^{\infty} \frac{u^j t^j}{j!} p(t),$$

where the last inequality follows since in the range $[\theta, \infty]$, the integral is positive. Then, the calculation can be simplified as follows:

$$\sum_{t\in(-\infty,\theta)} \exp(ut)p(t) \leq \sum_{t\in(-\infty,\infty)} (p(t) + utp(t))$$

$$+ \sum_{t\in(-\infty,\theta)} \sum_{j=2}^{\infty} \frac{u^j t^j}{j!} p(t)$$

$$\leq 1 + 0 + \sum_{t\in(-\infty,\theta)} \sum_{j=2}^{\infty} \frac{u^j t^j}{j!} p(t)$$

$$\leq 1 + \sum_{t\in(-\infty,\theta)} \frac{t^2}{\theta^2} \sum_{j=2}^{\infty} \frac{u^j \theta^j}{j!} p(t) \qquad \text{since } t < \theta \text{ in this range}$$

$$\leq 1 + \frac{1}{\theta^2}(\exp(u\theta) - 1 - u\theta)\mathrm{E}_r[Z_i^2].$$

For the second term, we have

$$\sum_{t>\theta} \exp(ut)p(t)$$

$$\leq \sum_{\ell=u\theta}^{\infty} \exp(\ell+1) \left(\Pr\left[Z_i > \frac{\ell}{u} + 1\right] - \Pr\left[Z_i > \frac{\ell}{u}\right]\right)$$

$$\leq \sum_{\ell=u\theta}^{\infty} \exp(\ell+1) \Pr\left[Z_i > \frac{\ell}{u} + 1\right]$$

$$\leq \sum_{\ell=u\theta}^{\infty} \exp(\ell+1) \Pr\left[Y_i^2 > \sigma_i^2 + \frac{\ell}{u} + 1\right]$$

$$\leq \sum_{\ell=u\theta}^{\infty} \exp(\ell+1) \exp\left(-\frac{\ell/u + \sigma_i^2}{2\sigma_i^2}\right)$$

$$\leq \sum_{\ell=u\theta}^{\infty} \exp\left(\ell + 1 - \frac{\ell}{2u\sigma_i^2} - \frac{1}{2}\right)$$

$$\leq \sqrt{e} \sum_{\ell=u\theta}^{\infty} \exp\left(\ell - \frac{\ell}{2u\sigma_i^2}\right).$$

By assumption of the lemma, since $u < \frac{1}{4\sigma_i^2}$, we have that $\ell - \frac{\ell}{2u\sigma_i^2} < -\frac{\ell}{4u\sigma_i^2}$. With this restriction,

$$\sum_{t>\theta} \exp(ut)p(t) \leq \sqrt{e} \cdot \sum_{\ell=u\theta}^{\infty} \exp\left(-\frac{\ell}{4u\sigma_i^2}\right)$$

$$\leq 2\sqrt{e} \cdot \exp\left(-\frac{u\theta}{4u\sigma_i^2}\right) = 2\sqrt{e} \cdot \exp\left(-\frac{\theta}{4\sigma_i^2}\right) \leq 4\exp\left(-\frac{\theta}{4\sigma_*^2}\right).$$

By putting together the two parts, we have that

$$\mathrm{E}_r[\exp(uZ_i)] \leq \left(1 + \frac{1}{\theta^2}(\exp(u\theta) - 1 - u\theta)\mathrm{E}_r[Z_i^2]\right)$$

$$+ 4\exp\left(-\frac{\theta}{4\sigma_*^2}\right).$$

Choosing $\theta = 4\sigma_*^2 \ln(k/\delta)$, the proof is complete. $\square$

## 9.3 Continued proof of Lemma 9

We finish the proof of Lemma 9 by showing that $g_s$ is decreasing. To this end, we prove that for all $A \subseteq [d]$ and for all $a \in [d] \setminus A$, $g_s(A \cup \{a\}) \leq g_s(A)$. Recalling the definition of $g_s(A)$, we have

$$g_s(A) = \mathrm{E}\left[\exp\left(u \sum_{i=1}^{s-1} \hat{Z}_i\right) \mid h^{-1}(s) = A\right]$$

$$= \mathrm{E}\left[\mathrm{E}\left[\exp\left(u \sum_{i=1}^{s-1} \hat{Z}_i\right) \mid \forall j : h(j)\right] \mid h^{-1}(s) = A\right], \tag{12}$$

where the inner expectation is over the random variables $\{r_j\}$ only. Since $h$ is completely independent we have that

$$g_s(A) = \sum_{\substack{(h_1,\ldots,h_d)\in[k]^d, \\ \forall j: h_j = s \Leftrightarrow j \in A}} \frac{\mathrm{E}\left[\exp\left(u\sum_{i=1}^{s-1} \hat{Z}_i\right) \mid \forall j : h(j) = h_j\right]}{(k-1)^{d-|A|}},$$

and similarly

$$g_s(A \cup \{a\})$$

$$= \sum_{\substack{(h_1,\ldots,h_d)\in[k]^d, \\ \forall j: h_j = s \Leftrightarrow j \in A\cup\{a\}}} \frac{\mathrm{E}\left[\exp\left(u\sum_{i=1}^{s-1} \hat{Z}_i\right) \mid \forall j : h(j) = h_j\right]}{(k-1)^{d-|A|-1}}.$$

Therefore it is sufficient to show that for all

$$(h_1,\ldots,h_{a-1},h_{a+1},\ldots,h_d) \in [k]^{d-1}$$

with $\forall j \neq a : h_j = s \Leftrightarrow j \in A$ it holds that

$$\sum_{h_a \in [k]\setminus\{s\}} \frac{\mathrm{E}\left[\exp\left(u\sum_{i=1}^{s-1} \hat{Z}_i\right) \mid \forall j : h(j) = h_j\right]}{k-1}$$

$$\geq \mathrm{E}\left[\exp\left(u\sum_{i=1}^{s-1} \hat{Z}_i\right) \mid \forall j \neq a : h(j) = h_j, h(a) = s\right].$$

We shall prove the following stronger inequality: for all $(h_1,\ldots,h_{a-1},h_{a+1},\ldots,h_d) \in [k]^{d-1}$ with $\forall j \neq a : h_j = s \Leftrightarrow j \in A$ and for all $h_a \in [k] \setminus \{s\}$ it holds that

$$\mathrm{E}\left[\exp\left(u\sum_{i=1}^{s-1} \hat{Z}_i\right) \mid \forall j : h(j) = h_j\right]$$

$$\geq \mathrm{E}\left[\exp\left(u\sum_{i=1}^{s-1} \hat{Z}_i\right) \mid \forall j \neq a : h(j) = h_j, h(a) = s\right].$$

Now observe that only $r$ are random in the above expectations and that $\hat{Z}_i$ are conditionally independent given $h$. Therefore,

$$\mathrm{E}\left[\exp\left(\sum_{i=1}^{s-1} u\hat{Z}_i\right) \mid h\right] = \prod_{i=1}^{s-1} \mathrm{E}\left[\exp\left(u\hat{Z}_i\right) \mid h\right].$$

From the non-negativity of the exponential function, it follows that it is sufficient to show that for all $i = 1, \ldots, s-1$ and for all $(h_1, \ldots, h_{a-1}, h_{a+1}, \ldots, h_d) \in [k]^{d-1}$ with $\forall j \neq a : h_j = s \Leftrightarrow j \in A$ and for all $h_a \in [k] \setminus \{s\}$ it holds that

$$E_L \geq E_R, \text{ where} \quad (13)$$
$$E_L = E\left[\exp\left(u\hat{Z}_i\right) \mid \forall j : h(j) = h_j\right]$$
$$E_R = E\left[\exp\left(u\hat{Z}_i\right) \mid \forall j \neq a : h(j) = h_j, h(a) = s\right].$$

We prove inequality (13) by a case analysis. If $h_a \neq i$, then $E_L = E_R$ by definition. If $h_a = i$ and the $i$th bucket of $E_L$'s hash function is bad, then $E_L = G(u, \sigma_*^4) \geq E_R$, as shown earlier in Corollary 8.

If $h_a = i$ and the $i$th bucket of $E_L$'s hash function is good then, the $i$th bucket of $E_R$'s hash function is also good. As before, define

$$V_a = \sum_{j \neq a} \sum_{g \neq a, j \neq g} r_j r_g x_j x_g \delta_{h(j)i} \delta_{h(g)i},$$

and

$$W_a = \sum_{j \neq a} r_j x_j x_a \delta_{h(j)i} \delta_{h(a)i}.$$

Again, note that if the $i$th bucket is good as assumed then $\hat{Z}_i = Z_i = V_a + r_a W_a$. Therefore we have that

$$E_L = E[E\left[\exp\left(uV_a + u \cdot r_a W_a\right) \mid \forall j : h(j) = h_j, \forall j \neq a : r_j\right]$$
$$\mid \forall j : h(j) = h_j]. \quad (14)$$

Now observe that only $r_a$ is random in the inner expectation and

$$E\left[uV_a + u \cdot r_a W_a \mid \forall j : h(j) = h_j, \forall j \neq a : r_j\right] = uV_a.$$

Thus from $E[\exp(x)] \geq \exp(E[x])$, it follows that

$$E\left[\exp\left(uV_a + u \cdot r_a W_a\right) \mid \forall j : h(j) = h_j, \forall j \neq a : r_j\right] \geq \exp\left(uV_a\right)$$

as before. Plugging the latter into (14) we arrive at

$$E_L \geq E\left[\exp\left(uV_a\right) \mid \forall j : h(j) = h_j\right] \quad (15)$$
$$= E\left[\exp\left(uV_a\right) \mid \forall j \neq a : h(j) = h_j, h(a) = s\right].$$

Here the last equality follows from the fact that all $r$ and $h$ values are independent and since $V_a$ does not depend on $a$. If $h^{-1}(s) = A \cup \{a\}$ and the $i$ hash bucket is good as assumed then $\hat{Z}_i = Z_i = V_a$ and we observe that

$$E\left[\exp\left(uV_a\right) \mid \forall j \neq a : h(j) = h_j, h(a) = s\right] = E_R \quad (16)$$

Combining (15) and (16), we conclude that $E_L \geq E_R$ for all cases and hence $g_s$ is decreasing as claimed.

## 9.4 Proof of Theorem 11 (i)

PROOF. Recall that the random variable $\hat{Z}_i$ is defined as

$$\hat{Z}_i = \begin{cases} Z_i & \text{if } \mathcal{G}_i, \\ \frac{1}{u} \log G(u, \sigma_*^4) & \text{else.} \end{cases}$$

Note that $G(u, \sigma_*^4) > 1$, and hence for $u > 0$, $\frac{1}{u} \log G(u, \sigma_*^4) > 0$. Also recall that $\mathcal{G}_i$ is the indicator vector for bucket $i$ being good, and $\mathcal{G}$ is the indicator for the hash function being good. By definition of $\hat{Z}_i$, since $\frac{1}{u} \log G(u, \sigma_*^4) > 0$, $\sum_i Z_i \wedge \mathcal{G} < \sum_i \hat{Z}_i$ and hence we have that,

$$\Pr\left[\sum_i Z_i > \epsilon\right] \leq \Pr\left[\sum_i Z_i > \epsilon \wedge \mathcal{G}\right] + \Pr[\bar{\mathcal{G}}]$$
$$\leq \Pr\left[\sum_i \hat{Z}_i > \epsilon\right] + \delta. \quad (17)$$

Thus, we prove a bound on $E[\exp(u \sum_i \hat{Z}_i)]$ and thus bound on $\Pr[\sum_i \hat{Z}_i > \epsilon]$. Taking expectations over both $r$ and $h$, and using $\Pr[\bar{\mathcal{G}}] < \delta$,

$$E[\exp(u\hat{Z}_i)] \leq E[\exp(uZ_i) \mid \mathcal{G}] \Pr[\mathcal{G}] + G(u, \sigma_*^4) \Pr[\bar{\mathcal{G}}]$$
$$\leq 1 + \frac{1}{\theta^2}(\exp(u\theta) - 1 - u\theta)\left(\frac{1}{k^2}(1 - \delta) + \sigma_*^4 \delta\right) + \frac{4\delta}{k},$$

where we combined the appropriate terms from the two parts of the sum. Recall that $\sigma_*^4 < 1$. By choosing $\delta < \frac{1}{k^2}$, we have that

$$E[\exp(u\hat{Z}_i)] \leq 1 + \frac{1}{\theta^2}(\exp(u\theta) - 1 - u\theta)\frac{2}{k^2} + \frac{4\delta}{k}$$
$$\leq \exp\left(\frac{2}{k^2\theta^2}(\exp(u\theta) - 1 - u\theta) + \frac{4\delta}{k}\right).$$

Taking the product over the $k$ terms, by using Lemma 10,

$$E[\prod_i \exp(u\hat{Z}_i)] \leq \prod_i E[\exp(u\hat{Z}_i)]$$
$$\leq \exp\left(\frac{2}{k\theta^2}(\exp(u\theta) - 1 - u\theta) + 4\delta\right).$$

For completeness, we show how to determine the optimal $u$.

$$\Pr\left[\sum_i \hat{Z}_i > \epsilon\right] \leq \frac{E[\exp(u \sum_i \hat{Z}_i)]}{\exp(u\epsilon)} \leq \prod_i E[\exp(u\hat{Z}_i)] \exp(-u\epsilon)$$
$$\leq \exp\left(\frac{2}{k\theta^2}(\exp(u\theta) - 1 - u\theta) + \delta - u\epsilon\right) \leq \exp\left(H(u) + \delta\right),$$

where we define $H(u) = \frac{2}{k\theta^2}(\exp(u\theta) - 1 - u\theta) - u\epsilon$. Setting the derivative $H'(u) = 0$, we get $\frac{2}{k\theta^2}(\theta \exp(u\theta) - \theta) - \epsilon = 0$, and hence

$$u = \frac{1}{\theta} \ln\left(1 + \frac{k\epsilon\theta}{2}\right) = \frac{1}{4\sigma_*^2 \ln(k/\delta)} \ln\left(1 + \frac{4k\epsilon\sigma_*^2 \ln(k/\delta)}{2}\right).$$

Note that we need to restrict $u < \frac{1}{4\sigma_*^2}$. We need $\frac{\ln(1 + \frac{4k\epsilon\sigma_*^2 \ln(k/\delta)}{2})}{\ln(k/\delta)} < 1$, which is true if setting $\frac{4k\epsilon\sigma_*^2 \ln(k/\delta)}{2} < \frac{k}{2\delta}$, or $\epsilon < \frac{k}{4\delta(1+\alpha)\ln(k/\delta)}$, which is permissive. Using this value of $u$, we have that (skipping the simplifications)

$$H = \frac{2}{k\theta^2}\left(\frac{k\epsilon\theta}{2} - \left(1 + \frac{k\epsilon\theta}{2}\right)\ln\left(1 + \frac{k\epsilon\theta}{2}\right)\right)$$
$$\leq \frac{2}{k\theta^2} \frac{\frac{k^2\epsilon^2\theta^2}{4}}{2 + \frac{2k\epsilon\theta}{6}} \leq \frac{-k\epsilon^2}{2(2 + \frac{k\epsilon\theta}{3})},$$

which is the trick that Bernstein uses: $(1 + x)\ln(1 + x) - x \leq \frac{-x^2}{2 + 2x/3}$. Plugging in this value of $H$, we get that

$$\Pr\left[\sum_i \hat{Z}_i > \epsilon\right] \leq \exp\left(\frac{-3k\epsilon^2}{4(3 + (1 + \alpha)\epsilon \ln(k/\delta))} + 4\delta\right). \quad \square$$

The proof of Theorem 11(ii) is similar to the above and is omitted in this version.